\documentclass[aps,prb,twocolumn]{revtex4}
\usepackage{amsmath}
\usepackage{amssymb}
\usepackage{graphicx}
\usepackage{dcolumn}
\usepackage{bm}

\def\be{\begin{equation}}
\def\ee{\end{equation}}
\def\bea{\begin{eqnarray}}
\def\eea{\end{eqnarray}}

\begin{document}

\title{Phonons under confinement: Finite Hilbert space, coarse-grained $U(1)$ charge and a new transport mechanism}
\author{Yuan Gao and K.A. Muttalib}
\affiliation{Department of Physics, University of Florida, P.O. Box 118440, Gainesville, FL 32611-8440}

\today

\begin{abstract}
An important feature of solid-state or cold atom systems in low dimensions is the restricted oscillations of ionic/atomic degrees of freedom in the confining directions, for which the conventional phonon from canonical quantization is not an ideal description. In this work we propose a general recipe to introduce this feature to otherwise unrestricted systems by mapping displacement fields to spin degrees of freedom, dubbed “soft-corization". In the strong-confinement limit $S=1/2$, we demonstrate the validity of the approach with a  harmonic chain; the results lead to a gapped fermionic model showing the absence of acoustic modes as a signature of the confinement of the entire chain. Despite the non-conserved total particle number in the presence of nearest-neighbor parings, the model holds a coarse-grained $U(1)$ symmetry with conserved “polarization charge" in the absence of nearest-neighbor hopping. This enables a new transport mechanism via "dimer-tumbling" instead of single-particle hopping. We then introduce a short-range interaction via anharmonicity; for energy scales slightly above the gap, it leads to a (quantum) phase transition to a nonreciprocal state carrying dimer current and heat current with spontaneous time reversal symmetry breaking of the type $\hat{T}^2=+1$. The non-reciprocity can be probed by measuring the change of inductance due to gauge fields in and out of the ordered phase.

\end{abstract}
\maketitle

\newpage

\section{Introduction} 
During the past several decades, low dimensional systems advanced our understanding of condensed matter in various ways. On one hand, topology starts to play an important role in classifying matter \cite{KKJ16, K86, CLM23, QZ11}, in addition to Landau’s symmetry related paradigms. On the other hand, spatial confinements can lead to breakdown of Landau’s Fermi liquid description \cite{H81, Z25, Z26}. While these issues were addressed widely for electronic systems, their fundamental impact on the phononic systems, to our knowledge, has been discussed only for in-plane motions of ionic/atomic degrees of freedom \cite{J25, HMHM18}.

In free-standing low-dimensional materials or nano-structures, the confining boundaries are free to shift in space with the interior of the system. As a result, the confining potential is usually modeled with various boundary conditions \cite{BW98,Za25,STM17,K16}, or through explicit modification of phonon wavefunctions \cite{RWL81,CF86,L06,S14}, and generally allows acoustic modes of out-of-plane vibrations. However, for more commonly used clamped experimental setup such as low-dimensional materials on a substrate or cold atoms in optical lattice, the spatial confinement of the ions/atoms at sample-host interface restricts the perpendicular motion of all interior particles as well, costing finite energy for the whole system to move away from the host. Thus, the corresponding transverse modes should be optical or gapped. Although it is possible to introduce a gap to the dispersion in a particular microscopic phononic model \cite{AG13}, the generic canonical quantization procedure with an infinite Hilbert space fails to capture the qualitative confinement on the vibrational amplitude of each atom (more details in Section II). Thus, conventional phonon description breaks down.

In this work, we propose a general approach dubbed {\it soft-corization} to introduce spatial confinement in an otherwise unconfined model by mapping displacement fields and its canonical momenta to spin degrees of freedom, reducing it to a finite dimensional Hilbert space. While the framework remains suitable for arbitrary spin allowing an effective low-energy theory, our goal is to show, with a simple proof-of-concept example, that the approach works and can lead to non-trivial observable features. Towards that goal we start with a 1D harmonic chain with one acoustic and one optical band and choose the simplest limit $S=1/2$. While this effectively becomes a hard core limit of our soft-corization framework, it will nevertheless allow us to explore many of the consequences of the spatial confinement  without going through the complexities of a larger spin model.
   
Given an effective spin-1/2 model, we apply a Jordan-Wigner transformation to convert it to fermionic degrees of freedom that yields a phonon model with non-conserving total fermion number. The resulting gapped band structure is reminiscent of a CDW model \cite{FP21, A22, BH20}, which is a special case of the Rice-Mele model \cite{Mc23, CF21, XCN10}, but with slightly different discrete symmetries. While the ground state is not a condensation, it looks similar to a BCS wavefunction. The low energy sector of the model could be identified as a massive Dirac fermion and could be bosonized to a sine-Gordon model with a relevant mass term. This shows directly the effectiveness of our approach in introducing a gap in a bosonic model as a result of spatial confinement.

Despite the explicit breaking of the global $U(1)$ symmetry by pairing correlations between nearest-neighbor sites, the model still holds a coarse-grained continuous symmetry when treated as a bipartite system in the absence of nearest-neighbor hopping correlations. The conserved quantity, dubbed {\it polarization charge}, is related to the difference in particle number between sublattices and enforces that in ground state only dimers exist. Including anharmonicity in the model is equivalent to adding interactions between these fermionic degrees of freedom. We show that such interactions can lead to novel phases of matter. The nearest neighbor density-density interactions from a trivial form of anharmonicity can lead to nontrivial instability of the normal state within a thin energy shell above the gap. This can give rise to an imbalance in the one-body backscattering in the Dirac fermion picture which we call  {\it nonreciprocity}; it is associated with broken time reversal symmetry of $T^2=+1$, distinct from the TR symmetry protecting Kramer's degeneracy in helical liquids \cite{WBZ06} with $T^2=-1$ which is explicitly broken by the mass gap in the model.
With periodic boundary condition, emergent dimer current and heat current are present in the time-reversal symmetry-breaking (TRSB) phase and are proportional to the nonreciprocity,  which plays the role of an order parameter. In real space, the TRSB states feature residual staggered dimer creation/annihilation resulting in the tumbling of dimers. This coarse-grained $U(1)$ symmetry allows coupling to (artificial) gauge fields \cite{DGJO11} in a new way. The order parameter also enters the paramagnetic current as equilibrium response to gauge fields at zero temperature, and thus can be probed.

The paper is organized as follows: After the introduction in Section I, Section II introduces the notion of soft-corization, the mapping for the spatial confinement of a harmonic chain.  In Section III, we focus on the strong confinement limit and provide detailed  consequences in several subsections.  The effective Hamiltonian is derived in subsection A and the associated symmetries are discussed in subsection B. Subsection C includes properties of ground state and excitations, while polarization charge and heat currents are discussed in subsection D which illustrates the emergence of the non-reciprocal state. Subsection E introduces the non-reciprocity order-parameter while subsection F discusses the coupling to a gauge-field that can probe the ordered state. Finally, subsection G discusses the paramagnetic current in and out of the non-reciprocal phase. Section IV includes a summary and outlook. Additional details are given in Appendices A-D.


\section{Soft-corization} 

Our goal is to spatially confine the individual ions/atoms at the surface/interface in the out-of-plane direction. We show that the bonds between these degrees of freedom and the rest of the system result in collective modes distinct from conventional phonons. 

We recall that an isolated harmonic oscillator located at lattice site $j$ can be described by the Hamiltonian 
\bea
H_j= \pi^{2}_{j}+\omega^{2}u^{2}_{j}=2\omega( \hat{n}_j+1/2), 
\eea
where $u_j$ is the out-of-plane displacement from its equilibrium position and $\pi_j$ is the canonical momentum, which satisfy $[u_{j},\ \pi_{k}]=i\delta_{jk}$. $\hat{n}_j=a^{+}_j a_j$ is the number operator of bosons satisfying $[ a_j,\ a^{+}_j]=1$. The amplitude of the displacement reads $\Delta u_{j,n}=\sqrt{<n|u^{2}_{j}|n>}=\sqrt{ (\hat{n}_j+1/2)/\omega}$ and reveals the fact that the energy level or the boson occupation number dictates the amplitude of the out-of-plane motion. However, for real scalar bosons it is not possible to restrict the occupation number at an arbitrary temperature as required by the spatial confinement. This implies that conventional boson is not the appropriate degree of freedom to be used in this context.

What we need is a restricted Hilbert space of finite dimensions, so that even at infinite temperature the occupation number is still confined. Spin operators $S_j$ with $[S^{-}_j,\ S^{+}_k]=-2S^{z}_j\delta_{jk}$, expressed in terms of bosons $b_j$ defined as $S^-_j=b_j$, $S_j^{\dag}=b_j^{\dag}$ and $S_j^z=b_j^{\dag}b_j-S$ where S is the maximum spin \cite{MM56} naturally meet this need. This leads to the definition of soft-core bosons
 \bea
  [b_{j},\ b^{+}_{k}]=2(S-n_{j}) \delta_{jk} 
\eea 
and allows the boson occupation numbers to vary from 0 to $2S$ on a given site and still preserve the symmetric permutation rule between different sites. We call the mapping from displacements to these soft-core bosons {\it soft-corization}. For half-integer $S$, the soft-corization can be explicitly implemented in the following way:
\bea
 u_{j} = \frac{ b^{+}_{j}+b_{j}}{2\sqrt{S-n_{j}}\sqrt{\omega}},\; \pi_{j}=i\frac{b^{+}_{j}-b_{j}}{2\sqrt{S-n_{j}}}\sqrt{\omega};\ S>n_{j},\cr
 u_{j} = i\frac{b^{+}_{j}-b_{j}}{2\sqrt{n_{j}-S}\sqrt{\omega}}, \; \pi_{j}=\frac{ b^{+}_{j}+b_{j}}{2\sqrt{n_{j}-S}}\sqrt{\omega}; \; S<n_{j} .
\eea

The next step is to rewrite the Hamiltonian of the to-be-confined system in terms of soft-core bosons. In this way the occupation number is confined without having to choose any explicit form of the confining potential that makes the system intrinsically anharmonic. The recipe can then generate, within a universal framework, effective theories that capture the essential qualitative features of confined systems where only lowest several energy levels are allowed. 


\section{Strong-confinement limit } 

For the remainder of this work, we will limit ourselves to a one-dimensional (1D) harmonic chain to show that even within a simple proof-of-concept example, the soft-corization approach reveals several non-trivial consequences of the spatial confinement including different symmetries. 
We will focus on the limit of strong confinement with $S=1/2$; the only allowed boson occupation numbers in this case are either 0 or 1, effectively corresponding to hard-core bosons. Thus, for vacant sites  $[b_{j},\ b^{+}_{j}]=+1$, while for an occupied site $[b_{j},\ b^{+}_{j}]=-1$.

For any Hamiltonian with only two-site correlations $H=\sum_{j,k} [H_{j}+ H_{jk}]$, one needs to figure out all two-site matrix elements for all possible occupation configurations. For example, $H^{pq}_{jk}$ stands for the matrix element between site $j$ with occupation number $p$ and site $k$ with occupation number $q$. Combining them in the following way guaranties that for a given configuration only the corresponding matrix elements survive and leads to the effective Hamiltonian
\bea
H^{eff}=\sum_{jk}\frac{1}{2}[\{ \hat{n}_{j}, H^{1}_{j}\}+\{(1-\hat{n}_{j}), H^{0}_{j}\}\cr
+\{\hat{n}_{j} \hat{n}_{k}, H^{11}_{jk}\}+\{(1-\hat{n}_{j})(1-\hat{n}_{k}), H^{00}_{jk}\}\cr
+\{\hat{n}_{j}(1-\hat{n}_{k}), H^{10}_{jk}\}+\{(1-\hat{n}_{j})\hat{n}_{k}, H^{01}_{jk}\}],
\label{heff}
\eea
where anti-commutators are to make the effective Hamiltonian Hermitian.


\subsection{Effective Hamiltonian } 

Hamiltonian for a 1D harmonic chain reads
\bea
\sum_{j}\ \hat{H}_{j}= \sum_{j}\ \tilde{h}\pi^{2}_{j}+ \tilde{J}\omega^{2}( u^{2}_{j}-u_{j}u_{j+1}). 
\label{hc}
\eea
Without any confinement, there is an optical flat band with $\xi_{2}=\tilde{h}$, and one acoustic branch with dispersion $\xi_{1k}=\tilde{J}(1-\cos k)$ for lattice spacing set to $1$, which features gapless excitations. 

When the system is confined, it is legitimate to relax $\tilde{J}$ to be occupation number dependent, e.g. the coupling strength between two occupied sites is slightly off from that between unoccupied cites. With the two-site matrix elements given in Appendix A and eq.~(\ref{heff}), the effective model in terms of the Pauli spin operators now reads
\bea
\hat{H} &=& \frac{\omega}{4}\sum_{j}\ -2 (\tilde{h}+ \tilde{J}_{0}) \sigma^{z}_{j}  \cr
&-& \frac{\tilde{J}_{11} -\tilde{J}_{00}}{2}  (\sigma^{y}_{j}\sigma^{y}_{j+1}- \sigma^{x}_{j}\sigma^{x}_{j+1}).
\eea
Here $\tilde{J}_{0}$ is the local potential for a vacant site, $\tilde{J}_{11}$ is the coupling constant between 2 occupied sites, and $\tilde{J}_{00}$ is the coupling constant between 2 vacant sites. We note that despite the absence of global $U(1)$ symmetry with conserved total spin $\sum_{j}\ \sigma^{z}_{j}$, there still exists a continuous symmetry with conserved staggered spin $\sum_{j}\ (-)^{j}\sigma^{z}_{j}$, which reveals the bipartite nature of the system. After redefinition of coefficients and a Jordan-Wigner transformation, the Hamiltonian can be rewritten as
\bea
\hat{H} &=& \sum^{N}_{j}\ 2h c^+_{j} c_{j} + \frac{J}{2} (c^{+}_{j-1} c^{+}_{j} + c^{+}_{j} c^{+}_{j+1} + h.c.),\cr
h &=& -(\tilde{h}+ \tilde{J}_{0})\omega/4; \;\;\; J=-( \tilde{J}_{11}-\tilde{J}_{00})\omega/8 .
\eea
Clearly, while the total particle number is uncertain, for each site the upper limit is one. This is the strong-confinement limit.


\subsection{Symmetries  }

The notion of global continuous symmetry in lattice systems is tricky; fields are only defined on discrete sites so that transformations are never truly global. Despite the fact that there is no global continuous symmetry on the original lattice, formally coarse-graining it to a bipartite lattice renders possible the existence of global (for all doubled unit cells) continuous symmetry of the resulting sublattice spinor. Treating the system as two sublattices, with odd sites ($A$) and even sites ($B$) respectively, the absence of nearest neighbor hopping terms gives rise to a global $U(1)$ symmetry for the Nambu spinor of the bipartite system, $ c_j=(c_{A,j} \ c^+_{B,j})^T $. If we rewrite the Hamiltonian in terms of bipartite fermions
\bea
\hat{H} = \sum^{\frac{N}{2}}_{j}\ 2h c^+_{A,j} c_{A,j} + 2h c^+_{B,j} c_{B,j} \cr
+J (c^+_{A,j} c^+_{B,j} + c^+_{B,j} c^+_{A,j+1} + h.c.),
\label{Hr}
\eea
where $N$ is total number of sites, a phase rotation of the Nambu spinor $c_j \rightarrow e^{i\phi} c_j$ is a symmetry operation, and the total number of A-particles and B-holes, dubbed {\it polarization charge}, is conserved:
\bea
\left[\sum^{\frac{N}{2}}_{j}(\ \hat{n}_{A,j}-\hat{n}_{B,j}+1),\ \hat{H}\right]=0.
\eea 

This model shares similar band structure with the CDW model, which is a special class of Rice-Mele model with uniform nearest neighbor hopping but staggered local potential, whose $U(1)$ symmetry is defined for spinor $ c^{CDW}_j=(c_{A,j} \ c_{B,j})^T $ and conserves total fermion number. We note that since the $U(1)$ symmetry holds in the CDW model even without coarse-graining of the lattice, the two models show distinct behavior when coupled to gauge fields.

In Fourier space, the Hamiltonian becomes
\bea
\hat{H} = \sum_{-\frac{\pi}{2}<k<\frac{\pi}{2}} c^{+}_{k} H(k) c_{k},\;\;\; c_{k}=(c_{A,k} \ c^{+}_{B,-k})^T \cr 
H(k) = 2J \sin k \;\sigma_2+2h\sigma_3 . \;\;\;\;\;\;\;\;
\label{Hofk}
\eea
There are several discrete symmetries associated with this Hamiltonian. A time-reversal symmetry transforming the spinor as $\hat{T} c_k=\sigma_0 \hat{K} c_{-k}$ satisfying $\hat{T}^2=1$ transforms $H(k)$ as $\sigma_0 H(k)\sigma_0 =H^*(-k)$. This symmetry operation doesn’t rotate the spinor, thus works on the two species $A$ and $B$ individually. 

In addition, the non-symmorphic \cite{A22, BH20, TM24} inversion and charge conjugation operations $P c_k=\sigma_3 c_{-k}$ and $C c_k = \sigma_1c_{-k}^{\dag}$ transform $H(k)$ as   
$\sigma_3 H(k) \sigma_3 =H(-k)$ and 
$\sigma_1 H(k) \sigma_1 =-H^*(-k)$, respectively.

Phonons in the strong-confinement limit can now be re-bosonized, made possible by the massive Dirac fermion behavior at long distances. One can always make the dispersive terms diagonal by applying a global rotation,
\bea
U&=&R_y(\frac{\pi}{2}) R_z(\frac{\pi}{2})= e^{-i\frac{\sigma_y}{2}\frac{\pi}{2}} e^{-i\frac{\sigma_z}{2}\frac{\pi}{2}}, \cr
c^D_{k} &=& (c_{R,k} \ c_{L,k})^T= U (c_{A,k} \ c^+_{B,-k})^T, \cr
H^{D}(k)&=&U H(k) U^{+}=2J \sin k\; \sigma_3 + 2h\sigma_1 
\label{HD}
\eea
where $R_i$ is the rotation gate around i-axis and $c_{L,k}$ and $c_{R,k}$ represent the left and right moving states. With the halved Brillouin zone from $-\frac{\pi}{2}$ to $\frac{\pi}{2}$, the transformed spinor contains two oppositely moving modes. In this language, the broken continuous symmetry conserving total number of $A$ and $B$ sublattices becomes the broken continuous chiral symmetry, while the coarse-grained $U(1)$ becomes the $U(1)$ symmetry conserving the total number of left and right movers. For small $k$, this is a massive Dirac fermion with fermi velocity $2J$ and mass $2h$, which could be bosonized \cite{C75, FG97} into a sine-Gordon model with Lagrangian density,
\bea
L(t,x)=\frac{1}{2\pi} [\frac{1}{2J} (\partial_t \phi)^2-2J(\partial_x \phi)^2]-4h\cos(2\phi).
\eea
Thus, at long distances, the original harmonic chain is mapped to a continuum phonon model with a relevant gap, which confirms the spatial confinement.


\subsection{Ground state and excitations}
To solve the model (\ref{Hofk}), one can define a canonical (Bogoliubov) transformation $U_k$ to diagonalize the Hamiltonian $H(k)$:
\bea
U_k = \left(
\begin{array}{cc} u_k & -v^*_k \cr
v_k & u^*_k 
\end{array} \right)
\eea
so that 
\bea
H^{’}(k) &=& U_k H(k) U^+_k= E_{k+} \sigma_3 \cr 
\gamma_k&=&(\gamma_{k+} \ \gamma_{k-})^T = U_k c_{k} \cr
 E_{k\pm} &=&\pm 2\sqrt{h^2+J^2\sin^{2}k} .
 \eea
 The unitarity of $U_k$ leads to
$|u_k|^2+|v_k|^2=1$,
$|u_k|^2-|v_k|^2=\frac{2h}{ E_{k+}}$ and 
$-2u_k v_k =\frac{i2J\sin k}{ E_{k+}}$.
The $\gamma_k$ particles have 0 chemical potential, and two bands are gapped by $4h$. The ground state is just where the whole lower band is filled and the upper band is empty,
\bea
|G\rangle=\Pi_k (u_k+v^*_k c^+_{A,k}c^+_{B,-k}) |vac\rangle.
\eea
Note that this is not the BCS wavefunction, since the form of the wavefunction depends on the gauge choice of the Nambu spinor, which implies that it is not a condensate state. For each $k$, the number of polarization charge carried by the ground state is 1, which indicates that on average, each doubled unit cell holds one localized polarization charge (dimer-vacuum mixture).

Excitations need to be excitonic, since $\gamma_k$ has $U(1)$ symmetry inherited from the sublattice Nambu spinor, similar to the situation for singlet superconductors where there is continuous symmetry conserving total spin along the quantization axis. In excited states, dimers are broken and the system becomes a conductor of quasi particles.


\subsection{Polarization charge and heat currents}
Since the conserved charge is properly defined, for the rest of this work, we explore the behavior of polarization charge and heat current at zero temperature. Both charge and heat currents can be defined using the continuity equation
\bea
\partial_t \hat{Q}_j +\partial_j \hat{J}_j =0.
\eea
In lattice systems, the gradient term reduces to $\hat{J}_{j} - \hat{J}_{j-1}$, and using Heisenberg equation of motion, it can be rewritten as,
\bea
\hat{J}_j - \hat{J}_{j-1} = i [\hat{Q}_j,\ \hat{H}_{j-1}+\hat{H}_{j}+\hat{H}_{j+1}],
\eea
where only nearest neighbor interactions have been assumed to exist. 

For polarization charge, $\hat{Q}_j = \hat{n}_{A,j} -\hat{n}_{B,j}+1$ and the charge current reads
\bea
\hat{J}^c_j = -iJ (c^+_{B,j} c^+_{A,j+1}- h.c.).
\label{jcr}
\eea
For heat, $\hat{Q}_j = \hat{H}_j$ and the heat current reads
\bea
\hat{J}^h_j =  i[2hJ(c^+_{A,j} c^+_{B,j} + c^+_{B,j} c^+_{A,j+1} - h.c.)\cr
- \frac{J^2}{2}( c^+_{A,j} c_{A,j+1} + c^+_{B,j} c_{B,j+1} - h.c.)].
\eea
We define current density
\bea
\bar{\hat{ J}}= (\frac{N}{2})^{-1} \sum_j \ \hat{J}_j = \int\frac{dk}{2\pi} \hat{J}(k).
\eea
The heat carrying part of the charge current, in terms of either the sublattice modes or the left and right moving states, can be written as 
\bea
\bar{\hat{ J}}^d &=& \int\frac{dk}{2\pi} (-)J \sin k (c^{+}_{A,k} c^{+}_{B,-k}+h.c.) \cr 
&=&  \int\frac{dk}{2\pi} i J\sin k(c^{+}_{R,k} c_{L,k}-h.c.).
\label{jcm}
\eea
Accordingly, the heat current density corresponding to this contribution is
$\bar{\hat{ J}}^h= -4h \bar{\hat{ J}}^d$, which shows that this heat current is carried by the dimer component of the polarization charge, and $\bar{\hat{ J}}^d$ is thus called dimer current. 

Following eq.~(\ref{jcm}), charge current density expectation value in any current carrying state can be rewritten as 
\bea
\langle \bar{\hat{ J}}^c\rangle =(\frac{N}{2})^{-1} \sum_j (-i)J (\langle c_{A,j} c_{B,j}\rangle- h.c.).
\eea
Comparing with eq.~(\ref{jcr}), the mechanism of the transport becomes apparent; charge is transported by annihilating a dimer, and then creating a new dimer involving the $B$ site of the former one, effectively {\it dimer tumbling} using $B$ sites as pivots, distinct from single-particle hopping. This new transport mechanism features physical shifting of individual $A$ particles across two lattice spacing, which is compatible with next-nearest neighbor hopping process. In momentum space, this can be understood as an imbalance in the two backscattering directions, which makes the system {\it nonreciprocal}.

We will show in the following subsections that the current carrying states can be achieved by spontaneous TRSB in the presence of anharmonicities, or by coupling to gauge fields given the coarse-grained $U(1)$ symmetry.


\subsection{Emergent currents from TRSB and nonreciprocal phase}
In this sub-section we show that in the presence of certain interactions, it is possible to have TRSB nonreciprocal states where $ \langle \bar{\hat{ J}}^d \rangle \neq 0$. We derive the free energy using path integral formalism and identify the boundary between reciprocal and nonreciprocal phases at zero temperature. We also obtain the expressions for currents by solving the order parameter self-consistently.

Interactions between fermions could arise from anharmonicity added to the harmonic chain (\ref{hc}). For example, $H^{int}=\tilde{V} \omega^3 \sum_j \ u^{2}_{j} u^{2}_{j+1}$, upon soft-corization and Jordan-Wigner transformation, becomes nearest neighbor density-density interaction of the form
\bea
\hat{ H}^{int} &=& \frac{V}{N} \sum_j \ c^{+}_{A,j} c^{+}_{B,j} c_{B,j} c_{A,j} + c^{+}_{B,j} c^{+}_{A,j+1} c_{A,j+1} c_{B,j}\cr
&=& \frac{4V}{N^2} \sum_{k,k^{’},q} \ c^{+}_{A,k} c^{+}_{B, k^{’}} \cos q \ c_{B, k^{’}-q} c_{A,k+q},
\eea
plus an additional local potential so that $h$ is modified to $h^{a}=h-V/2$, where $V=\omega \tilde{V}$. Allowing only pairs with zero total momentum (as with Cooper pairs) and keeping only the odd-parity triplet terms, the interaction part of the mean-field Hamiltonian can be written as 
\bea
\bar{\hat{H}}^{int} &=& \int \ \frac{dk}{2\pi} \tilde{\Delta} \sin k \;c_{B,-k} c_{A, k} + h.c. -\frac{|\tilde{\Delta}|^2}{V}, \cr
 \tilde{\Delta} &\equiv &V\left\langle \int \ \frac{dk}{2\pi} \sin k\; c^{+}_{A,k} c^{+}_{B, -k}\right\rangle.
 \label{OP}
 \eea
 Note that Im$\tilde{\Delta} $ is a purely renormalization effect of $J$ and doesn't break any symmetries. Now it is clear that $\Delta\equiv$ Re$\tilde{\Delta}$ is the {\it nonreciprocity  order parameter} which is proportional to emergent currents, and $\Delta \sin k \sigma_1$ is the TR breaking term added to (\ref{Hofk}) generating a mean field Hamiltonian
\bea
&&H_{mean}(k)=\Delta \sin k \sigma_1 + 2J \sin k \sigma_2 + 2h^{a}\sigma_3,\cr
&&\sigma_0 H_{mean}(k) \sigma_0 \neq  H_{mean}^*(-k).
\eea
where the renormalization of $h$ as discussed above is included.

The partition function of the mean field Hamiltonian reads
\bea
Z &=& \int D\Delta D\psi D\bar{\psi}  \exp(\Delta^2/V)  \cr
&\times & \exp \left[\frac{1}{\beta} \sum_m \int \frac{dk}{2\pi} \bar{\psi}_{\omega_m,k} (-)L(\omega_m,k) \psi_{\omega_m,k}\right],
\eea
where $\int D…$ represents a functional integral, $\omega_m $ is Matsubara frequency, the spinor is defined as $ \psi_{\omega_m,k} = ( \psi_{A,\omega_m,k} \ \bar{\psi}_{B,-\omega_m,-k})^T$, and Lagrangian density $ L(\omega_m,k)= -i\omega_m\sigma_0+ H_{mean}(k)$.

Integrating out the fermions, we are able to identify the free energy in terms of the order parameter,
\bea
F = &-&\frac{\Delta^2}{V} - \frac{1}{\beta} \sum_m \int \frac{dk}{2\pi} \ln \det(L(\omega_m,k))\cr
\det(L(\omega_m,k)) &=& (i\omega_m)^2+E_{k+}E_{k-} -\Delta^2 \sin^{2}k . 
\eea
Around the phase boundary where the order parameter is small, we Taylor expand the free energy to fourth order of $\Delta$ at low temperatures,
\bea
\delta F = \Delta^2 \left(-\frac{1}{V} + \int \frac{d\omega}{2\pi} \frac{dk}{2\pi} \frac{ \sin^{2}k }{(i\omega_m)^2+E_{k+}E_{k-}}\right)\cr
+ \Delta^4 \frac{1}{2} \int \frac{d\omega}{2\pi} \frac{dk}{2\pi} \frac{ \sin^{4}k }{((i\omega_m)^2+E_{k+}E_{k-})^2}.
\eea
As long as the fourth order term is always positive and the second order term changes sign at the phase boundary, the symmetry-breaking state is at least metastable. 

\begin{figure*}
\includegraphics[angle=0,width=0.35\textwidth]{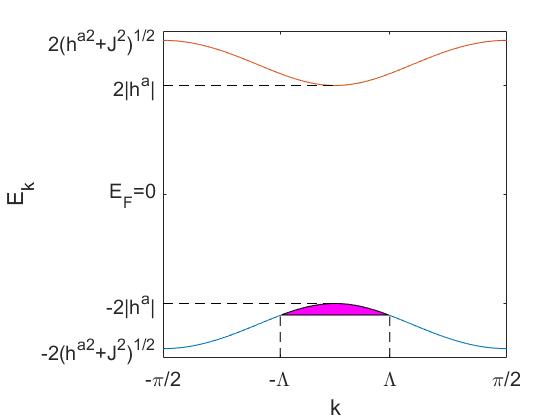}
\includegraphics[angle=0,width=0.35\textwidth]{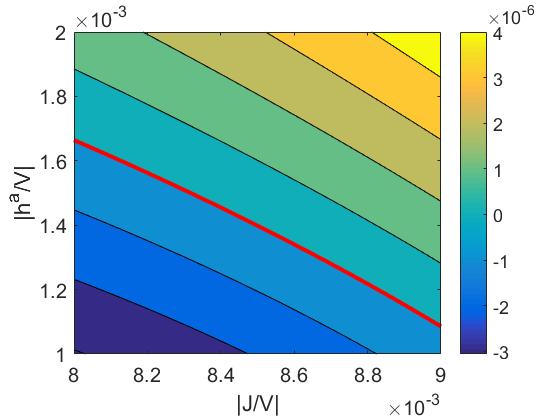}
\caption{Left: Dispersion relation of the model (\ref{Hofk}), with shaded unstable momentum shell in the effective theory. Right: Phase diagram with $|h^{a}/V|$, $|J^{}/V|$ as control parameters, for $\Lambda=0.5$. Red line is the boundary between reciprocal (positive values) and non-reciprocal phase (negative values).}
\label{dispersion}
\end{figure*}

The phase boundary is identified by solving $ \frac{1}{V} + \Pi=0$, which indicates the instability of the scattering process characterized by the bubble 
\bea
\Pi= -\int \frac{d\omega}{2\pi} \frac{dk}{2\pi} \frac{ sin^{2}k }{(i\omega_m)^2+E_{k+}E_{k-}}.
\label{bubble}
\eea

The equation only has solution for $V<0$. As shown in Appendix B and Figure \ref{dispersion} (Left), the dispersion relations and the corresponding density of states suggest an effective theory within the range $- \Lambda <k<\Lambda $, or equivalently $ -2|h^b|< E_{-}<-2|h^{a}|$ where $h^b=\sqrt{h^{a2}+\Lambda^2 J^2}$. With appropriate change of variables and setting  $T=0$ we get,
\bea
16\pi|J^{’}|^3= \int^{-2|h^{a’}|}_{ -2|h^{b'}|} dE^{’}_{-} \sqrt{E^{’2}_{-}-4h^{a’2}},\;\;\; x^{’}=\frac{x}{-V} .
\eea
Figure \ref{dispersion} (Right) shows that in general, nonreciprocal phase exists when $|V|\gg |h^{a}|, |J|$, which might seem unrealistic. However, $ h^{a}$ and $J$ are not the original energy scales in the problem; in terms of original energy scales, the nonreciprocal phase emerges when $|V|\gg |\tilde{h}+ \tilde{J}_{0}+2V|, |\tilde{J}_{11}-\tilde{J}_{00}|$, which become realistic for $V < 0$. In addition, the quartic term of the free energy is positive as needed.

\begin{figure*}
\includegraphics[angle=0,width=0.45\textwidth]{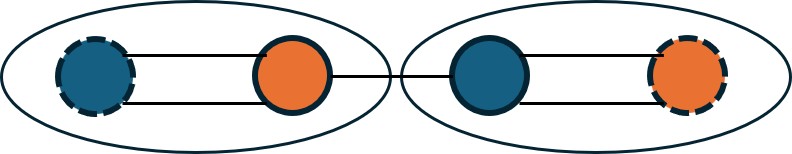}\;\;\;\;\;\;\;\;\;\;\;\;
\includegraphics[angle=0,width=0.45\textwidth]{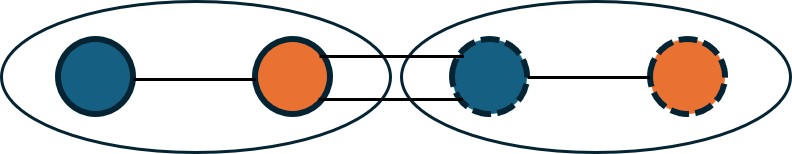}
\caption{Nonreciprocal states configurations. Blue dots are $A$ sites, and orange dots are $B$ sites. Solid boundary of the dots mean occupied, while dashed boundaries mean unoccupied. Double lines represent creation bonding and single lines represent annihilation bonding. Left: Only intra-cell dimer creation and inter-cell dimer annihilation exist, effectively dimers tumble to left. Right: Only inter-cell dimer creation and intra-cell dimer annihilation exist, effectively dimers tumble to right. }
\label{ssb}
\end{figure*}

The order parameter 
\bea
\Delta
= V\int \ \frac{dk}{2\pi} \sin k \; {\rm Re}\langle c^{+}_{A,k} c^{+}_{B, -k} \rangle
\label{orderp}
\eea
is to be evaluated self-consistently, where the expectation value taken with the symmetry-breaking state (eigenstate of $H_{mean}(k)$) will yield its own $\Delta$ dependence. As shown in Appendix C, this leads to,
\bea
16\pi \left(J^{’2}+\frac{\Delta^{’2}}{4}\right)^{\frac{3}{2}}= \int^{-2|h^{a’}|}_{ -2|h^{b'}|} d\epsilon ^{’}_{-} \sqrt{\epsilon ^{’2}_{-}-4h^{a’2}} \cr
\eea
where $\epsilon_{\pm}$ is eigenenergy of $H_{mean}(k)$ and $T=0$ is taken. The two-fold degenerate ground states feature opposite directions of dimer and heat currents, as illustrated in Figure \ref{ssb}. The emergent current densities are then given by
\bea
\langle \bar{\hat{ J}}^d\rangle =-2VJ^{'}  \Delta^{'} ,\;\;\; \langle\bar{\hat{ J}}^h\rangle =-8V^2 h^{a’}J^{'} \Delta^{'} .
\eea

Note that without external fields, Bloch theorem \cite{W19, KS19} forbids currents for open boundary condition and restricts the magnitude of charge currents to $O(N^{-1})$ for periodic boundary condition. However, the forbidden current is the full polarization charge current; the emergent dimer current on the other hand is only the heat carrying part of the total current. Unfortunately, it is challenging to probe the heat current directly by, e.g., coupling to a temperature gradient; this is because thermal fluctuations at any finite temperature will destroy the order in 1D \cite{MW66, Y24}.

\subsection{Coupling to gauge field}

In the presence of gauge fields, currents can exist at zero temperature both in and out of nonreciprocal phase. However, the system is charge-neutral and does not couple to an electromagnetic field. Thus, an ideal probe of the system involves exploiting the fact that the system has the coarse-grained $U(1)$ symmetry, and it  can be coupled in principle to an artificial gauge field, in a way that leaves the coupling term invariant under gauge transformation:
\bea
(c_{A,j} \ c^+_{B,j})^T \rightarrow e^{i\phi_j} (c_{A,j} \ c^+_{B,j})^T,\cr\cr
{\cal{A}}(j,t) \rightarrow {\cal{A}}(j,t)+\partial_{j}\phi_j={\cal{A}}(j,t)+(\phi_{j+1}-\phi_j).
\eea
Going back to the real space effective Hamiltonian (\ref{Hr}), it is easy to check that the following coupling preserves this gauge invariance,
\bea
\sum^{\frac{N}{2}}_{j}\ \hat{H}^{’}_{j} =\sum^{\frac{N}{2}}_{j}\ 2h (\hat{n}_{A,j} +\hat{n}_{B,j})\cr
+J (c^+_{A,j} c^+_{B,j} + c^+_{B,j} c^+_{A,j+1} \exp(i{\cal{A}}(j,t)) + h.c.).
\eea
As expected, the gauge field resides only on the links between unit cells, which makes the system highly nonuniform even in long-wavelength limit. It could be anticipated that the system can develop unusual response to gauge field. By definition, the charge current in the presence of the gauge field is,
\bea
\hat{J}^{c’}_j = - \frac{\partial{\hat{H}^{’}_{j} }}{\partial{{\cal{A}}(j,t)}}\cr
= -iJ (c^+_{B,j} c^+_{A,j+1} \exp(i{\cal{A}}(j,t)) - h.c.).
\eea
Same result is obtained using the continuity equations.


\subsection{Paramagnetic currents in and out of the nonreciprocal phase}
In the very limited scope of this work, we study the equilibrium effects brought by the gauge field, the non-dissipative inductive currents. After detailed derivation shown in Appendix D, the first order inductive current in static and long-wavelength limit at zero temperature takes the form,
\bea
<\bar{\hat{ J}}_{ind}>= J^2 {\cal{A}} \int\frac{dk}{2\pi} \frac{1-2 sin^{2}k }{-\epsilon_{k-} }= K{\cal{A}},
\eea
where $ K>0$ is the paramagnetic susceptibility of the response.

The susceptibility depends on $\Delta$ through $\epsilon_{k-} $. As a result, the inductance of the system could be used as a probe of the nonreciprocal order. The change in the inductance upon entering the ordered phase is,
\bea
\delta K &=& K(\Delta)-K(0),\cr
K(\Delta)&=& \frac{J^2}{2\pi (J^2+\Delta^2/4)^{1/2}}
\times \left ( \int^{-2|h^a|}_{-2|h^b|}  \frac{d\epsilon_{-}}{\sqrt{\epsilon^2_{-} -4h^{a2}}}\right.\cr
&-&\left. \frac{1}{2 (J^2+\Delta^2/4)} \int^{-2|h^a|}_{-2|h^b|} d\epsilon_{-} \sqrt{\epsilon^2_{-} -4h^{a2}}\right).
\eea

Note that the introduction of gauge field explicitly breaks time reversal symmetry, and the current operator itself as the order parameter enters the coupling term. This makes the phase structure more complex in nonreciprocal phase, analogous to ferromagnetism in the presence of external magnetic field. Here we assume that the first-order transitions are avoided by choosing the parameters appropriately.


\section{Summary and outlook}
We established a systematic framework to effectively impose spatial confinements in the out-of-plane direction on an otherwise unrestricted low dimensional phononic system. This is achieved primarily by mapping the displacement field of individual atomic degrees of freedom to spin operators, which intrinsically restricts the Hilbert space to finite dimensions. For spin-1/2, the resulting gapped excitation for phonons in a 1D harmonic chain shows the existence of suppressed collective motion of ions/atoms. 

This effective spin-model holds a coarse-grained $U(1)$ symmetry and can exhibit exotic phases of matter in the presence of interactions. At the mean field level, an instability to a TRSB state could be achieved by controlling the effective local potential $h^{a}$, the nearest neighbor coupling strength $J$ as well as temperature in the presence of finite attractive interactions. We only considered $T=0$ case to avoid thermal fluctuations which are destructive to any order in 1D. In the TRSB phase, a finite non-dissipative heat current density and a corresponding dimer current density emerge. This nonreciprocal effect could be probed by measuring the change of paramagnetic inductance of the system as a response to an (artificial) gauge field. 

In this work, we only explored one specific interaction derived from one particular anharmonicity, while the fascinating complexity of anharmonicities could give rise to other exotic phases of matter. For example, with the help of forward and backscattering of the left and right movers the mass gap could become irrelevant in the low energy sector, allowing the system to enter a critical phase where various algebraic orders could survive. Alternatively, the existence of long-range interactions could protect the nonreciprocal order at finite temperatures where the emergent heat current could be coupled to temperature gradients \cite{GM24, GM25} and thus be probed more easily.

Our discussions have been limited to 1D, and we only mapped displacement fields to spin-1/2 operators. The generalization to 2D and higher spin is straightforward. The interplay between this generalized soft-core phonon and electrons, conventional phonons, magnons, etc. could make low dimensional systems an even more exciting platform for novel condensed matter phenomena. 
In particular, general soft-core phonons might be a good candidate for the localized scatterer \cite{MA17} that absorbs propagating phonons and reduces thermal conductivity significantly in silicon nanowires with surface roughness \cite{HCD08}, allowing for better thermoelectric devices.

\section{Acknowledgements}
We acknowledge helpful discussions with Yuxuan Wang.

\appendix


\section{Two-site matrix elements for effective Hamiltonian}
Here we lay out all two-site matrix elements needed for the construction of confined effective theory for (\ref{hc}).
\bea
\pi^{1}_{j}\pi^{1}_{j}=\frac{\omega}{2}(b^{+}_{j}b^{+}_{j}+b^{+}_{j}b_{j}+H.C.+[b_{j}, b^{+}_{j}])\cr
=\frac{\omega}{2}(4(S^{x}_{j})^{2}-1)=0,\cr
\omega^{2}u^{1}_{j}u^{1}_{j}=\frac{\omega}{2}(-b^{+}_{j}b^{+}_{j}+b^{+}_{j}b_{j}+H.C.+[b_{j}, b^{+}_{j}])\cr
=\frac{\omega}{2}(4(S^{y}_{j})^{2}-1)=0,\cr
H^{1}_{j}=\tilde{h}\pi^{1}_{j}\pi^{1}_{j}+\tilde{J}_{1}\omega^{2}u^{1}_{j}u^{1}_{j}=0,\\
\pi^{0}_{j}\pi^{0}_{j}=\frac{\omega}{2}(-b^{+}_{j}b^{+}_{j}+b^{+}_{j}b_{j}+H.C.+[b_{j}, b^{+}_{j}])\cr
=\frac{\omega}{2}(4(S^{y}_{j})^{2}+1)=\omega,\cr
\omega^{2}u^{0}_{j}u^{0}_{j}=\frac{\omega}{2}(b^{+}_{j}b^{+}_{j}+b^{+}_{j}b_{j}+H.C.+[b_{j}, b^{+}_{j}])\cr
=\frac{\omega}{2}(4(S^{x}_{j})^{2}+1)=\omega,\cr
H^{0}_{j}=\tilde{h}\pi^{0}_{j}\pi^{0}_{j}+\tilde{J}_{0}\omega^{2}u^{0}_{j}u^{0}_{j}\cr
=(\tilde{h}+\tilde{J}_{0})\omega,\\
\omega^{2}u^{1}_{j}u^{1}_{j+1}=\frac{\omega}{2}(-b^{+}_{j}b^{+}_{j+1}+b^{+}_{j}b_{j+1}+H.C.)\cr
=2\omega S^{y}_{j}S^{y}_{j+1},\cr
H^{11}_{jj+1}=-\tilde{J}_{11}\omega^{2}u^{1}_{j}u^{1}_{j+1}\cr
=-2 \tilde{J}_{11}\omega S^{y}_{j}S^{y}_{j+1},\\
\omega^{2}u^{0}_{j}u^{0}_{j+1}=\frac{\omega}{2}(b^{+}_{j}b^{+}_{j+1}+b^{+}_{j}b_{j+1}+H.C.)\cr
=2\omega S^{x}_{j}S^{x}_{j+1},\cr
H^{00}_{jj+1}=-\tilde{J}_{00}\omega^{2}u^{0}_{j}u^{0}_{j+1}\cr
=-2 \tilde{J}_{00}\omega S^{x}_{j}S^{x}_{j+1},\\
\omega^{2}u^{0}_{j}u^{1}_{j+1}=i\frac{\omega}{2}(b^{+}_{j}b^{+}_{j+1}-b^{+}_{j}b_{j+1}-H.C.)\cr
=-2\omega S^{x}_{j}S^{y}_{j+1},\cr
H^{01}_{jj+1}=-\tilde{J}_{01}\omega^{2}u^{0}_{j}u^{1}_{j+1}\cr
=2 \tilde{J}_{01}\omega S^{x}_{j}S^{y}_{j+1},\\
\omega^{2}u^{1}_{j}u^{0}_{j+1}=i\frac{\omega}{2}(b^{+}_{j}b^{+}_{j+1}+b^{+}_{j}b_{j+1}-H.C.)\cr
=-2\omega S^{y}_{j}S^{x}_{j+1}.\cr
H^{10}_{jj+1}=-\tilde{J}_{10}\omega^{2}u^{1}_{j}u^{0}_{j+1}\cr
=2 \tilde{J}_{10}\omega S^{y}_{j}S^{x}_{j+1}.
\eea


\section{Evaluating the bubble}
The bubble $\Pi$ in eq.~({\ref{bubble}) can be expressed in terms of Green’s functions of quasiparticles,
\bea
\Pi &=& \int \frac{d\omega}{2\pi} \frac{dk}{2\pi} \frac{ \sin k }{i\omega - E_{k-}} \frac{\sin k }{-i\omega - E_{-k-}}\cr
&=& \int \frac{d\omega}{2\pi} \frac{dk}{2\pi} \sin^{2} k \;G_{-}(\omega,k) \;G_{-}(-\omega,-k),
\eea
where $G_{-}$ is the Green’s function of the lower-band quasiparticles. There are other ways to interpret the process by including upper-band quasiparticles, but we are only interested in zero temperature where thermal fluctuations are absent. The bubble implies that ``dressed" polarization charges scatter with each other and suit themselves into TR breaking states with lower free energy in certain regions of parameter space.

The bubble can be rewritten as
\bea
\Pi &=& -\int \frac{dk}{2\pi} \frac{\sin^{2} k}{2 E_{k-}} \int \frac{d\omega}{2\pi} \left(\frac{1}{i\omega-E_{k-}} - \frac{1}{i\omega + E_{k-}}\right)\cr
&=& -\int \frac{dk}{2\pi} \frac{\sin^{2} k}{2 E_{k-}} (f(E_{k-})- f(E_{k+}))\cr
&=& \frac{1}{4J^2} \int \frac{dk}{2\pi} \frac{E^2_{k-}-4h^{a2}}{2 E_{k-}} \tanh\frac{ E_{k-}}{2T}.
\eea
The energy scale of the problem matters. The density of states of the free model is given by $N(E_{\pm}) =  \frac{1}{\pi} |dk/dE_{\pm}|=  \frac{1}{\pi} |\sqrt{h^{a2}+J^2\sin^{2}k}/(J^2\sin 2k) | $.
For small $k$, $ N(E_{\pm}) = |E_{\pm}|/2\pi|J| \sqrt{E^2_{\pm}-4h^{a2}}$ is dimensionless, which makes the four-fermion interaction marginal, while for energy scale much larger than the gap, $ N(E_{\pm})\sim E^{-1}_{\pm}$ makes the interaction relevant and unimportant at high energy. As a result, we only keep the effective theory within the range $- \Lambda <k<\Lambda $, or equivalently $ -2|h^b|< E_{-}<-2|h^{a}|$ where $h^b=\sqrt{h^{a2}+\Lambda^2 J^2}$ at zero temperature. A more careful treatment with RG is left for the future. Note that this cutoff is determined by the property of the free (non-interacting) model.

Within the thin momentum shell above the gap,
\bea
\Pi &=&  \frac{1}{8\pi|J|^3}  \int dE_{-} \frac{-E_{-}}{ \sqrt{E^2_{-}-4h^{a2}}} \frac{E^2_{-}-4h^{a2}}{2 E_{-}} \tanh\frac{ E_{-}}{2T}\cr
&=& -\frac{1}{16\pi|J|^3} \int^{-2|h^a|}_{-2|h^b|} dE_{-} \sqrt{E^2_{-}-4h^{a2}} \tanh\frac{ E_{-}}{2T}
\eea
which is positive. Thus, attractive interaction is needed for a phase transition.


\section{Self-consistency equation for the order parameter}
We recall the definition (\ref{orderp}):
\bea
\Delta  = V\;\int  \frac{dk}{2\pi} \sin k \;{\rm Re} \langle c^{+}_{A,k} c^{+}_{B, -k}\rangle .
\eea
The expectation value taken with the symmetry-breaking state (eigenstate of $H_{mean}(k)$) will yield $\Delta$ dependence so that the order parameter could be solved self-consistently:
\bea
\langle c^{+}_{A,k} c^{+}_{B,-k} \rangle 
= -u_{k}v_{k}(f(\epsilon_{k+})- f(\epsilon_{k-}))\cr
=\frac{\Delta \sin k+i2J\sin k}{2| \epsilon_{k}|}(f( \epsilon_{k+})- f(\epsilon_{k-})),
\eea
where $\epsilon_{k\pm}=\pm 2 \sqrt{h^2+(J^2+\Delta^2/4)\sin^{2}k}$ is the dispersion of the mean field Hamiltonian. With the modified density of states $ N( \epsilon_{\pm}) = \frac{1}{2\pi\sqrt{ J^2+\Delta^2/4}} \frac{|\epsilon _{\pm}|}{\sqrt{\epsilon ^2_{\pm}-4h^{a2}}}$,
\bea
 \Delta  &=& \frac{\Delta \ V}{16\pi (J^2+\Delta^2/4)^{3/2}} \cr 
&\times & \int^{-2|h^a|}_{-2\sqrt{h^{a2}+ \Lambda^2 (J^2+ \Delta^2/4)}} d\epsilon_{-} \sqrt{\epsilon^2_{-} -4h^{a2}}\cr
&\times& \tanh\frac{ \epsilon_{-} }{2T}.
\eea
Assuming both $\Lambda $ and $\Delta $ to be small, one can drop $\Delta$ in the lower bound of the integral and finally
\bea
&&\frac{16\pi(J^{2}+\Delta^{2}/4)^{3/2}}{V} \cr
&=& \int^{-2|h^{a}|}_{ -2|h^{b}|} d\epsilon_{-} \sqrt{\epsilon ^{2}_{-}-4h^{a2}} \tanh\frac{ \epsilon_{-} }{2T}.
\eea


\section{Inductive currents} 

Since the transport currents are generally absent in gapped systems in the linear response regime (no tunneling), we don’t employ Kubo formula here for a detailed calculation but perform an equilibrium perturbation theory so that only inductive currents are kept. We expand the phase factor to first order of the gauge field,
\bea
\hat{H}^{’}_{j} &=& \hat{H}^{0}_{j} - {\cal{A}} (j,t) \hat{J}^{c0}_j,\cr
\cr
\hat{J}^{c’}_j &=& \hat{J}^{c0}_j + {\cal{A}} [J(c^+_{B,j} c^+_{A,j+1} + h.c.)],\cr
\cr
\langle\hat{J}^{c’}_j\rangle &=& \langle\hat{J}^{c0}_j \rangle_{’} + {\cal{A}} \langle J(c^+_{B,j} c^+_{A,j+1} + h.c.)\rangle_{0},
\eea
where $\langle…\rangle_{0}$ denotes expectation values taken in states $\rho_0$ satisfying $[\rho_0,\ \sum^{\frac{N}{2}}_{j}\hat{H}^{0}_{j}] =0$ in the absence of the gauge field, and $\langle…\rangle_{’}$ denotes expectation values taken in states $\rho^{’}$ satisfying $[\rho^{’},\ \sum^{\frac{N}{2}}_{j}\hat{H}^{’}_{j}] =0$ in the presence of the gauge field, which is generally invalid for time dependent perturbation unless nonequilibrium steady states are achieved. The first term in the perturbed charge current operator is paramagnetic and the second term is diamagnetic (although with a plus sign). Here the “turn on” of the gauge field is treated as adiabatic, so the current will be totally inductive. 

Taking the static and long-wavelength limit, the charge current density becomes,
\bea
\langle\bar{\hat{J}}^{c’}\rangle &=& 2J \left(- {\rm Re} \int\frac{dk}{2\pi} \sin k\; \langle c^{+}_{A,k} c^{+}_{B,-k}\rangle_{’}\right.\cr 
&-& \left. {\rm Im} \int\frac{dk}{2\pi} \cos k\;\langle c^{+}_{A,k} c^{+}_{B,-k}\rangle_{’} \right)\cr
&-&2J{\cal{A}} \left({\rm Re} \int\frac{dk}{2\pi} \cos k\; \langle c^{+}_{A,k} c^{+}_{B,-k}\rangle _{0} \right.\cr
&-& \left. {\rm Im} \int\frac{dk}{2\pi} \sin k\; \langle c^{+}_{A,k} c^{+}_{B,-k} \rangle_{0} \right).
\eea
where the perturbed Hamiltonian density 
\bea
H^{’}(k)&=& (J{\cal{A}}+\Delta)\sin k\sigma_{1}\cr
&+&J (2\sin k+{\cal{A}}\cos k) \sigma_{2} + 2h^{a}\sigma_{3}
\eea
 gives rise to 
 \bea
  \epsilon^{’}_{k\pm}=\pm\sqrt{\epsilon^2_{k\pm}+{\cal{A}}^2 J^2+{\cal{A}}(2J^2\sin 2k+2J\Delta \sin^{2} k)}.\cr
   \eea
  At $T=0$, the expression reduces to, 
\bea
\langle \bar{\hat{J}}^{c’}\rangle &=& J^2 {\cal{A}} \int\frac{dk}{2\pi} \frac{1}{-\epsilon^{’}_{k-}} -J^2 {\cal{A}} \int\frac{dk}{2\pi} \frac{2 \sin^{2} k }{-\epsilon_{k-}}\cr
&+&J\Delta \int\frac{dk}{2\pi} \frac{ \sin^{2} k }{-\epsilon^{’}_{k-}}.
\label{charge}
\eea
For the first term, since it has explicit linear $\cal{A}$ in the front, we could drop the $\cal{A}$ dependence in $\epsilon^{’}_{k-}$, so that it reduces to $\epsilon_{k-}$ and we can combine the first two terms, thus we have the total inductive current in the main text.
The third term in (\ref{charge}) is the emergent dimer current in the presence of gauge field, where the dependence on $\cal{A}$ is negligibly small.

\end{document}